\documentclass[11pt,letterpaper]{article}
\pdfoutput=1

\usepackage[linesnumbered,boxruled,vlined]{algorithm2e}

\usepackage[margin=1in]{geometry}
\usepackage[T1]{fontenc}
\usepackage{microtype}
\usepackage{setspace}   

\usepackage{amsmath}
\usepackage{amsthm}
\usepackage{mathtools}
\usepackage{newpxtext,newpxmath}
\allowdisplaybreaks[2]

\usepackage{graphicx}
\graphicspath{{./}{./figures/}}
\usepackage{booktabs}
\usepackage{threeparttable}
\usepackage[margin=15pt,font=small,labelfont={bf}]{caption}
\usepackage{subcaption}
\usepackage{enumitem}
\setlist{itemsep=2pt, topsep=4pt, parsep=0pt}
\setlist[enumerate]{label=(\arabic*)}
\usepackage{comment}
\usepackage{cite}

\usepackage[dvipsnames]{xcolor}
\colorlet{LinkBlue}{blue!60!black}
\colorlet{CiteGreen}{green!45!black}
\colorlet{UrlViolet}{violet}
\usepackage[unicode,pdfencoding=auto,colorlinks=true]{hyperref}
\hypersetup{linkcolor=LinkBlue, citecolor=CiteGreen, urlcolor=UrlViolet}
\usepackage{bookmark}
\usepackage{thm-restate}
\usepackage[capitalise,noabbrev,nameinlink]{cleveref}

\theoremstyle{plain}
\newtheorem{theorem}{Theorem}
\newtheorem{lemma}[theorem]{Lemma}
\newtheorem{corollary}[theorem]{Corollary}

\newtheorem{definition}[theorem]{Definition}

\usepackage{tcolorbox}
\newtcolorbox{problemdef}{
    colback=white, colframe=black, boxrule=0.8pt,
    arc=0pt, outer arc=0pt,
    left=4pt, right=4pt, top=2pt, bottom=2pt,
    before skip=6pt, after skip=6pt
}

\newcommand{\dist}{\mathsf{dist}}        
\newcommand{\tail}{\mathsf{tail}}        
\newcommand{\head}{\mathsf{head}}        
\newcommand{\Eout}{E^{+}}                

\usepackage[whole]{bxcjkjatype}

\newcommand{\randorder}{\UseTextAccent{OMS}{\textcircled}{r}}
\newcommand{\randordernote}{The author ordering was randomized using \url{https://www.aeaweb.org/journals/policies/random-author-order/generator}. It is requested that citations of this work list the authors separated by \texttt{\textbackslash{}textcircled\{r\}} instead of commas.}

\title{A Simple Optimal Algorithm for Directed Multi-Failure Replacement Paths}
\author{
Kaito Harada \\
The University of Osaka\\
\texttt{k-harada@ist.osaka-u.ac.jp} \and
\randorder\thanks{\randordernote} \and
Tatsuya Terao \\
Kyoto University\\
\texttt{ttatsuya@kurims.kyoto-u.ac.jp}
}

\date{}

\begin{document}

\maketitle

\begin{abstract}
    Given a graph $G = (V, E)$ and two vertices $s, t \in V$, the \emph{$f$-fault replacement paths ($f$-FRP)} problem asks to compute, for every set $F$ of at most $f$ edges, the distance from $s$ to $t$ in $G \setminus F$.
    
    In this paper, we present a $3$-FRP algorithm in deterministic $O(n^3 \log n)$ time for directed weighted graphs.
    We also present a deterministic $O(n^4)$ time algorithm for $4$-FRP in directed weighted graphs, which implies a deterministic $O(n^f)$ time algorithm for $f$-FRP for every $f \ge 4$.
    Since the output size of $f$-FRP can be $\Theta(n^f)$, our algorithms are optimal for every $f \ge 4$ and nearly optimal for $f = 3$.
    Our algorithms also extend to the single-source and all-pairs settings with $f-1$ and $f-2$ failures, respectively, with the same time complexity as $f$-FRP.
    Moreover, our algorithms are remarkably simple.
\end{abstract}

\section{Introduction} \label{sec:intro}

The shortest path problem is one of the most fundamental problems in graph algorithms and theoretical computer science.
We study the \emph{replacement paths} problem, which concerns shortest paths in a failure-prone graph $G = (V, E)$ with $n = |V|$ and $m = |E|$.
The replacement paths problem is well studied in the literature~\cite{malik1989k,nardelli2001faster,bhosle2004replacement,hershberger2007difficulty,gotthilf2009improved,emek2010near,wulff2010solving,bernstein2010nearly,erickson2011computing,williams2011faster,RZ12,GV12,weimann2013replacement,lee2014replacement,ACC19,grandoni2019faster,CC19,chechik2020simplifying,CM20,GJM20,BCFS21,GPVX21,BCCFS22,CZ24,manoharan2024computing,chang2025optimal,CDWX25,NV26,HI26,HT26}, and it is practically motivated by its role as a basic model for fault-tolerant routing.
In the classical replacement paths (RP) problem, the number of failures is one.
Given two vertices $s, t \in V$, the RP problem asks to compute, for every edge $e \in E$, a shortest $s$-$t$ path that avoids $e$.

Vassilevska Williams, Woldeghebriel, and Xu~\cite{VWX22} initiated a study of the replacement paths problem under multiple edge failures.
The formal definition of the problem is as follows.

\begin{problemdef}
    \textbf{$f$-Fault Replacement Paths Problem ($f$-FRP)}

    \textbf{Input:}
    A graph $G = (V, E, w)$, vertices $s, t \in V$ and the number of failures $f$

    \textbf{Output:}
    The distance $\dist_{G \setminus F}(s, t)$ from $s$ to $t$ in $G \setminus F$ for each edge set $F \subseteq E$ with $|F| \le f$.
\end{problemdef}

Here, $1$-FRP is just the classical RP.
In RP, removing an edge outside a fixed shortest $s$-$t$ path does not change the $s$-$t$ distance, so only the at most $n-1$ edges on the path need to be considered.
In $f$-FRP, applying the same observation to each failure in turn shows that the $i$-th failed edge only needs to be chosen from a shortest $s$-$t$ path in the graph after the first $i-1$ failures.
Hence, it suffices to compute $\dist_{G \setminus F}(s, t)$ for only $O(n^f)$ edge sets $F$, and the output size of $f$-FRP can be $\Theta(n^f)$.
A natural question is how close the running time for $f$-FRP can get to the output size $n^f$.

Vassilevska Williams, Woldeghebriel, and Xu~\cite{VWX22} presented an $\tilde{O}(n^3)$ time\footnote{The $\tilde{O}(\cdot)$ notation hides polylogarithmic factors, i.e., $\tilde{O}(g(n)) = O(g(n)\,\mathrm{polylog}(n))$.} algorithm for $2$-FRP in directed weighted graphs.
Since $f$-FRP reduces to solving $(f-1)$-FRP on $G \setminus \{e\}$ for each of the at most $n-1$ edges $e$ on a shortest $s$-$t$ path, their algorithm yields an $\tilde{O}(n^{f+1})$ time algorithm for $f$-FRP for every $f \ge 2$.
Chi, Duan, Wang, and Xie~\cite{CDWX25} presented an $\tilde{O}(n^3)$ time algorithm for $3$-FRP in undirected weighted graphs, which implies an $\tilde{O}(n^f)$ time algorithm for $f$-FRP in undirected weighted graphs for every $f \ge 3$.

We also consider the single-source and all-pairs variants of $f$-FRP, where the distances are computed for every target $t \in V$ and for every pair $(s, t) \in V \times V$, respectively.
The formal definitions of the two problems are as follows.

\begin{problemdef}
    \textbf{$f$-Fault Single-Source Replacement Paths Problem ($f$-FSSRP)}

    \textbf{Input:}
    A graph $G = (V, E, w)$, a source $s \in V$ and the number of failures $f$

    \textbf{Output:}
    The distance $\dist_{G \setminus F}(s, t)$ from $s$ to $t$ in $G \setminus F$ for each target $t \in V$ and edge set $F \subseteq E$ with $|F| \le f$.
\end{problemdef}

\begin{problemdef}
    \textbf{$f$-Fault All-Pairs Replacement Paths Problem ($f$-FAPRP)}

    \textbf{Input:}
    A graph $G = (V, E, w)$ and the number of failures $f$

    \textbf{Output:}
    The distance $\dist_{G \setminus F}(s, t)$ from $s$ to $t$ in $G \setminus F$ for each $(s, t) \in V \times V$ and edge set $F \subseteq E$ with $|F| \le f$.
\end{problemdef}

Since the argument for $f$-FRP applies to each target $t$ and to each pair $(s, t)$, the output sizes of $f$-FSSRP and $f$-FAPRP can be $\Theta(n^{f+1})$ and $\Theta(n^{f+2})$, respectively.

For a single failure, $1$-FSSRP and $1$-FAPRP in directed weighted graphs can be solved in $\tilde{O}(mn + n^2)$ and $\tilde{O}(mn + n^3)$ time, respectively, by using the distance sensitivity oracle of Bernstein and Karger~\cite{BK09}.
Faster algorithms for $1$-FSSRP are known for unweighted graphs~\cite{CC19,CM20,GJM20,BCFS21,BCCFS22, DG22,HI26} and for graphs with small integer weights~\cite{GV12,grandoni2019faster,GPVX21}.
For two failures, Chi, Duan, Wang, and Xie~\cite{CDWX25} gave an $\tilde{O}(n^3)$ time algorithm for $2$-FSSRP in undirected weighted graphs.

\subsection{Our Results}

\Cref{tab:results} summarizes the best known running times for $f$-FRP, $f$-FSSRP, and $f$-FAPRP in weighted graphs, together with our new results.

\begin{table}[t]
    \centering
    \small
    \begin{threeparttable}
        \caption{Deterministic exact algorithms for weighted graphs with $f$ edge failures. Edge weights are positive reals for undirected graphs and arbitrary reals (no negative cycle) for directed graphs.
        Our algorithms work for directed graphs.} \label{tab:results}
        \begin{tabular}{@{}llllll@{}}
            \toprule
                    &          &             & \multicolumn{2}{c}{Previous results} & \\
            \cmidrule(lr){4-5}
            Problem & Failures & Output size & Undirected & Directed & \textbf{This paper} \\
            \midrule
            FRP   & $f = 1$   & $\Theta(n)$     & $O(m+n \log n)$\cite{malik1989k} & APSP-time\tnote{$\ddagger$}\;\cite{grandoni2019faster} & --- \\
                  & $f = 2$   & $\Theta(n^2)$     & APSP-time\tnote{$\ddagger$}\;\cite{NV26} & $O(n^3 \log^2 n)$\cite{VWX22} & $O(n^3)$ \\
                  & $f = 3$   & $\Theta(n^3)$     & $\tilde{O}(n^3)$~\cite{CDWX25}     & $\tilde{O}(n^4)$~\cite{VWX22}     & $O(n^3 \log n)$ \\
                  & $f \ge 4$ & $\Theta(n^f)$     & $\tilde{O}(n^f)$~\cite{CDWX25}     & $\tilde{O}(n^{f+1})$~\cite{VWX22} & $O(n^f)$ \\
            \midrule
            FSSRP & $f = 1$   & $\Theta(n^2)$     & APSP-time\tnote{$\ddagger$}\;\cite{grandoni2019faster}     & APSP-time\tnote{$\ddagger$}\;\cite{grandoni2019faster}     & --- \\
                  & $f = 2$   & $\Theta(n^3)$     & $\tilde{O}(n^3)$~\cite{CDWX25}     & $\tilde{O}(n^4)$\tnote{$\dagger$}\;\cite{VWX22}     & $O(n^3 \log n)$ \\
                  & $f \ge 3$ & $\Theta(n^{f+1})$ & $\tilde{O}(n^{f+1})$~\cite{CDWX25} & $\tilde{O}(n^{f+2})$\tnote{$\dagger$}\;\cite{VWX22} & $O(n^{f+1})$ \\
            \midrule
            FAPRP & $f = 1$   & $\Theta(n^3)$     & $O(n^3 \log^2 n)$~\cite{BK09}      & $O(n^3 \log^2 n)$~\cite{BK09}   & $O(n^3 \log n)$ \\
                  & $f \ge 2$ & $\Theta(n^{f+2})$ & $\tilde{O}(n^{f+2})$\tnote{$\dagger$}\;\cite{CDWX25} & $\tilde{O}(n^{f+3})$\tnote{$\dagger$}\;\cite{VWX22} & $O(n^{f+2})$ \\
            \bottomrule
        \end{tabular}
        \begin{tablenotes}
            \footnotesize
            \item[$\dagger$] Not stated explicitly; obtained by applying the cited algorithm for each target, each source, or each pair of vertices.
            \item[$\ddagger$] APSP-time denotes the running time to solve the all-pairs shortest paths (APSP) problem on $n$-vertex graphs. Following the breakthrough of Alman and Vassilevska Williams~\cite{alman2026trulysubquadratic3sumtruly}, the best known bounds are $n^{2.9945+o(1)}$ for polynomially bounded integer weights~\cite{Maj26}, and deterministic $O(n^{2.9997})$~\cite{CIMY26} and expected $O(n^{2.998})$~\cite{alman2026trulysubquadratic3sumtruly} for real weights.
        \end{tablenotes}
    \end{threeparttable}
\end{table}

In this paper, we develop a simple framework that solves $f$-FRP, $f$-FSSRP, and $f$-FAPRP on directed weighted graphs in nearly optimal time.
Our framework yields the following results.

\begin{theorem} \label{thm:2frp}
    The 2-FRP problem in directed real-weighted graphs can be solved in $O(n^3)$ time.
\end{theorem}

\begin{theorem} \label{thm:3frp}
    The 3-FRP, 2-FSSRP, and 1-FAPRP problems in directed real-weighted graphs can be solved in $O(n^3 \log n)$ time.
\end{theorem}

\begin{theorem} \label{thm:4frp}
    The 4-FRP, 3-FSSRP, and 2-FAPRP problems in directed real-weighted graphs can be solved in $O(n^4)$ time.
\end{theorem}

Since $f$-FRP reduces to $O(n)$ instances of $(f-1)$-FRP, and $f$-FSSRP and $f$-FAPRP reduce to $n$ instances of $f$-FRP and $f$-FSSRP, respectively, \cref{thm:4frp} implies the following corollary.

\begin{corollary}\label{cor:ffrp}
    For any integer $f \ge 4$, the $f$-FRP, $(f-1)$-FSSRP, and $(f-2)$-FAPRP problems in directed real-weighted graphs can be solved in $O(n^f)$ time.
\end{corollary}
We remark that, if given the failed edges, our framework is able to report an optimal replacement path $\pi$ in $O(|\pi|)$ time, where $|\pi|$ is the number of vertices in $\pi$.

Since the output size of $f$-FRP, $(f-1)$-FSSRP, and $(f-2)$-FAPRP can be $\Theta(n^f)$, the running time of our algorithm is optimal when $f \ge 4$.
Also, even when $f = 3$, our algorithm is nearly optimal.



\section{Preliminaries} \label{sec:prelimi}

Let $G = (V, E, w)$ be a simple directed graph with $n = |V|$ vertices and a weight function $w: E \to \mathbb{R}$.
Edge weights may be negative, but we assume that $G$ has no negative cycle.
We also assume without loss of generality that $n = 2^h$ for an integer $h \ge 0$, and we index the vertices as $V = \{v_0, v_1, \dots, v_{n-1}\}$; otherwise, we add fewer than $n$ isolated dummy vertices to $G$, which changes neither the distances between the original vertices nor the asymptotic running time.

For an edge $e = (u, v)$ we call $u$ the \emph{tail} and $v$ the \emph{head} of $e$, and denote them by $\tail(e)$ and $\head(e)$, respectively.
For a vertex $x$, let $\Eout(x)$ be the set of edges whose tail is $x$, and for a vertex set $X \subseteq V$, we let $\Eout(X) = \bigcup_{x \in X} \Eout(x)$.
For an edge set $F \subseteq E$, we denote by $G \setminus F$ the graph obtained from $G$ by removing the edges in $F$, and we write $\tail(F) = \{ \tail(e) \mid e \in F\}$.

The weight $w(P)$ of a walk or path $P$ is the sum of the weights of its edges.
For a graph $H$ and two vertices $u, v$ in $H$, the distance $\dist_H(u, v)$ is the minimum weight of a path from $u$ to $v$ in $H$, and $\dist_H(u, v) = \infty$ if $v$ is unreachable from $u$.

For a set $X$ and an element $x \in X$, we write $X \setminus x$ for $X \setminus \{x\}$.
For two directed paths or walks $\alpha$ and $\beta$ such that the last vertex of $\alpha$ is the first vertex of $\beta$, we denote their concatenation by $\alpha\circ\beta$.
We write $\langle x_0, x_1, \dots, x_k \rangle$ for the walk or path that visits the vertices $x_0, x_1, \dots, x_k$ in this order.

\section{Our Algorithm} \label{sec:algo}

We begin with an overview of our algorithm for $f$-FRP.
Our algorithm first constructs a data structure that returns $\dist_{G \setminus F}(s, t)$ for any failed edge set $F$ with $|F| \le f$, and then obtains the output by querying it.


The key structure is the \emph{$f$-sparsified shortcut graph} of a vertex set $X \subseteq V$ (\cref{subsec:shortcut}).
For every pair of vertices $x, y \in X$ and every edge $e$ leaving $x$, consider a shortest path in $G$ from $x$ to $y$ that starts with $e$ and has no internal vertices in $X$.
Roughly speaking, the $f$-sparsified shortcut graph stores only the $f+1$ shortest of these paths for each pair $x, y$, and hence it has only $O(f|X|^2)$ vertices and edges.
Suppose that $|F| \le f$ and every failed edge leaves a vertex of $X$.
Then, only the first edges of these paths can fail because the other edges leave vertices not in $X$.
Hence, for each pair $x, y$, at least one of the $f+1$ paths from $x$ to $y$ avoids $F$.
Based on this observation, we show that, after discarding the paths containing failed edges, the distance between any two vertices of $X$ in the $f$-sparsified shortcut graph is equal to the distance in $G \setminus F$.
Thus, using the $f$-sparsified shortcut graph of $X = \{s, t\} \cup \tail(F)$, we can compute $\dist_{G \setminus F}(s, t)$ in $O(f^6)$ time.

It remains to compute the $f$-sparsified shortcut graphs of all sets $X$ of this form in advance.
For $X = V$, the $f$-sparsified shortcut graph is essentially $G$ itself, and we obtain smaller sets by removing vertices from $X$ one at a time.
When a vertex $x$ is removed from $X$, paths may newly pass through $x$, and we update the graph in the same way as the update step of the Floyd--Warshall algorithm (\cref{subsec:removal}).
To compute all of these $f$-sparsified shortcut graphs efficiently, we organize the removals by recursively halving vertex sets (\cref{subsec:build}).

\subsection{Shortcut Graphs and Their Sparsification} \label{subsec:shortcut}

In this subsection, we introduce shortcut graphs and $f$-sparsified shortcut graphs.
For $X \subseteq V$, an $f$-sparsified shortcut graph $\mathcal{S}_f(X)$ has only $O(f|X|^2)$ vertices and edges.
Nevertheless, for every set $F \subseteq E$ with $|F| \le f$ and $\tail(F) \subseteq X$, the distance between any two vertices of $X$ in $\mathcal{S}_f(X) \setminus F$ is equal to the distance in $G \setminus F$.
In our algorithm, we use $f$-sparsified shortcut graphs to compute $\dist_{G \setminus F}(s, t)$ for each failed edge set $F \subseteq E$ with $|F| \le f$.

\begin{figure}[!tb]
    \centering
    \begin{subfigure}[b]{0.49\textwidth}
        \centering
        \includegraphics[width=\linewidth]{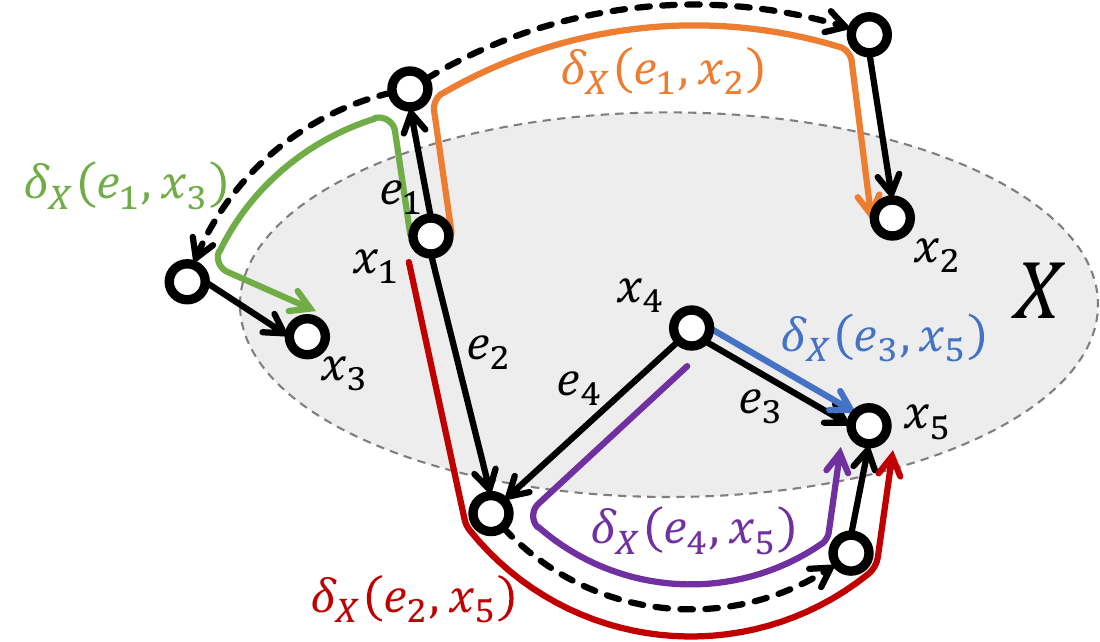}
        \caption{$X$-shortcut in $G$.}\label{fig:shortcut}
    \end{subfigure}\hfill
    \begin{subfigure}[b]{0.44\textwidth}
        \centering
        \includegraphics[width=\linewidth]{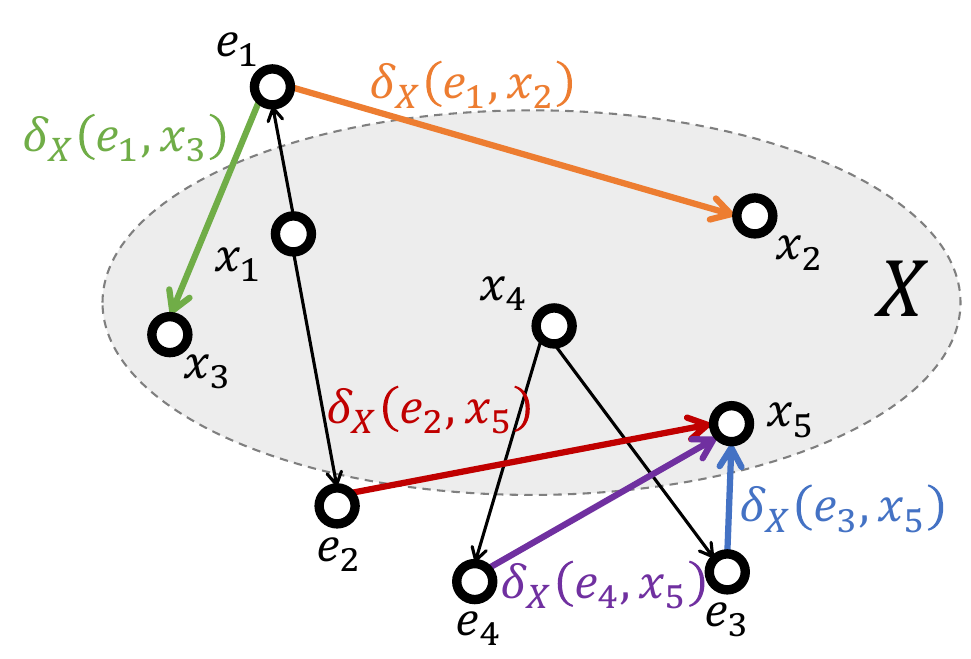}
        \caption{The shortcut graph $\mathcal{S}(X)$.}\label{fig:shortcut-graph}
    \end{subfigure}
    \caption{A graph $G$ with $X = \{x_1, x_2, \dots, x_5\}$ (shaded) and its shortcut graph $\mathcal{S}(X)$. (a) Solid black arrows are edges and dashed black arrows are paths outside $X$. Each colored path is an $X$-shortcut of weight $\delta_X(e_i, x_j)$. (b) Each edge $e_i$ of $G$ whose tail is in $X$ becomes a vertex of $\mathcal{S}(X)$. Black arrows are the departure edges. Colored arrows are the shortcut edges for the shortcuts of the same color in (a).}\label{fig:shortcut-figures}
\end{figure}

\paragraph*{Shortcut graphs.}
We first introduce \emph{shortcuts} and \emph{shortcut graphs} (see \cref{fig:shortcut-figures} for an example).

\begin{definition}[Shortcut]
    Let $X \subseteq V$ and let $x, y \in X$.
    An $X$-shortcut from $x$ to $y$ is a path from $x$ to $y$ in $G$, none of whose internal vertices belong to $X$.
    For an edge $e \in \Eout(x)$, we denote by $\delta_X(e, y)$ the shortest path length of an $X$-shortcut from $x$ to $y$ whose first edge is $e$ (and $\delta_X(e, y) = \infty$ if there is no such shortcut).
\end{definition}

\begin{definition}[Shortcut graph]
    For $X \subseteq V$, the shortcut graph $\mathcal{S}(X)$ is the weighted directed graph defined as follows:
    \begin{itemize}
        \item The vertex set is the disjoint union of $X$ and $\Eout(X)$.
            In other words, $\mathcal{S}(X)$ has one vertex for each vertex of $X$ and one vertex for each edge $e$ of $G$ with $\tail(e) \in X$.
            We denote the vertex corresponding to an edge $e$ also by $e$.
        \item For every $x \in X$ and $e \in \Eout(x)$, there is a departure edge $(x, e)$ of weight 0.
        \item For every $x, y \in X$ and $e \in \Eout(x)$, there is a shortcut edge $(e, y)$ of weight $\delta_X(e, y)$.
    \end{itemize}
\end{definition}

When we consider the shortcut graph $\mathcal{S}(X)$, we regard each edge $e \in E$ of $G$ with $\tail(e) \in X$ as a vertex of $\mathcal{S}(X)$, and we denote the vertex also by $e$.

We first give a recurrence for computing $\delta_X(e, y)$ in \cref{lem:removal}, which expresses $\delta_{X \setminus x}$ in terms of $\delta_X$ for a vertex $x \in X$.
A $V$-shortcut has no internal vertex, and hence $\delta_V(e, y) = w(e)$ if $\head(e) = y$ and $\delta_V(e, y) = \infty$ otherwise.
Starting from $\delta_V$ and removing the vertices of $V \setminus X$ one by one, we can compute $\delta_X(e, y)$ for any $X \subseteq V$ by dynamic programming.
Intuitively, the recurrence performs the same update as the Floyd--Warshall algorithm.

\begin{lemma}\label{lem:removal}
    Let $X \subseteq V$, $x \in X$.
    For each $u, v \in X \setminus x$ and each $e \in \Eout(u)$, 
    \[\delta_{X \setminus x}(e, v) = \min \left \{ \delta_X(e, v), \delta_X(e, x) + \min_{e' \in \Eout(x)} \delta_X(e', v) \right \}.\]
\end{lemma}

\begin{proof}
    We prove the equality by showing the two inequalities separately.

    $(\le)$ Every $X$-shortcut from $u$ to $v$ is also an $(X\setminus x)$-shortcut, so $\delta_{X \setminus x}(e, v) \le \delta_X(e, v)$.
    Moreover, let $\pi_1$ be an $X$-shortcut from $u$ to $x$ with first edge $e$, and let $\pi_2$ be an $X$-shortcut from $x$ to $v$.
    The internal vertices of the concatenation $\pi_1 \circ \pi_2$ are $x$ and the internal vertices of $\pi_1$ and $\pi_2$.
    Since $x \notin X \setminus x$ and the internal vertices of the $X$-shortcuts $\pi_1$ and $\pi_2$ are not in $X$, the walk $\pi_1 \circ \pi_2$ is an $(X \setminus x)$-shortcut from $u$ to $v$ with first edge $e$.
    Hence, we have $\delta_{X \setminus x}(e, v) \le w(\pi_1) + w(\pi_2)$.

    $(\ge)$ We may assume $\delta_{X \setminus x}(e, v) < \infty$.
    Let $\pi$ be an $(X \setminus x)$-shortcut from $u$ to $v$ with first edge $e$ and weight $\delta_{X \setminus x}(e, v)$.
    If $\pi$ does not contain $x$, then $\pi$ is an $X$-shortcut and $\delta_{X \setminus x}(e, v) = w(\pi) \ge \delta_X(e, v)$.
    Otherwise, $x$ is an internal vertex of $\pi$.
    Splitting $\pi$ at $x$ gives an $X$-shortcut $\pi_1$ from $u$ to $x$ with first edge $e$ and an $X$-shortcut $\pi_2$ from $x$ to $v$, so $\delta_{X\setminus x}(e, v) = w(\pi)= w(\pi_1) + w(\pi_2) \ge \delta_X(e, x) + \min_{e' \in \Eout(x)} \delta_X(e', v)$.
\end{proof}

Next, we show that, for every set $F$ of failed edges with $\tail(F) \subseteq X$, the distance between any two vertices of $X$ in $\mathcal{S}(X) \setminus F$ is equal to the distance in $G \setminus F$.
Here, $F$ is a set of edges of $G$, and each edge in $F$ corresponds to a vertex of $\mathcal{S}(X)$ since $\tail(F) \subseteq X$.
We denote by $\mathcal{S}(X) \setminus F$ the graph obtained from $\mathcal{S}(X)$ by removing the vertices corresponding to the edges in $F$.

\begin{lemma}\label{lem:preserve}
    Let $X \subseteq V$, let $x, y \in X$, and let $F \subseteq E$ be an edge set with $\tail(F) \subseteq X$.
    Then, $\dist_{G \setminus F}(x, y) = \dist_{\mathcal{S}(X) \setminus F}(x, y)$ holds.
\end{lemma}

\begin{proof}
    We prove the equality by showing the two inequalities separately.

    $(\ge)$ We may assume that $y$ is reachable from $x$ in $G \setminus F$.
    Let $\pi$ be a shortest path from $x$ to $y$ in $G \setminus F$.
    Let $x = z_0, z_1, \dots, z_k = y$ be the vertices of $X$ on $\pi$ in the order of appearance, and let $\pi_i$ be the subpath of $\pi$ from $z_{i-1}$ to $z_i$.
    Each $\pi_i$ is an $X$-shortcut, and its first edge $e_i \in \Eout(z_{i-1})$ is not in $F$.
    Hence $w(\pi_i) \ge \delta_X(e_i, z_i)$, and $\mathcal{S}(X) \setminus F$ contains the path $\langle z_{i-1}, e_i, z_i \rangle$ of weight $\delta_X(e_i, z_i)$.
    Concatenating these paths for $i=1, \dots, k$ gives a path from $x$ to $y$ in $\mathcal{S}(X) \setminus F$, and $\dist_{G \setminus F}(x, y) = w(\pi) = \sum_{i=1}^k w(\pi_i) \ge \sum_{i=1}^k \delta_X(e_i, z_i) \ge \dist_{\mathcal{S}(X) \setminus F}(x, y)$ holds.

    $(\le)$ Every path from $x$ to $y$ in $\mathcal{S}(X) \setminus F$ has the form $\langle z_0, e_1, z_1, e_2, \dots, e_k, z_k \rangle$ with $z_0 = x$, $z_k = y$, $e_i \in \Eout(z_{i-1}) \setminus F$, and weight $\sum_{i=1}^k \delta_X(e_i, z_i)$.
    For each $i$, let $\pi_i$ be an $X$-shortcut from $z_{i-1}$ to $z_i$ with first edge $e_i$ and weight $\delta_X(e_i, z_i)$.
    The first edge $e_i$ of $\pi_i$ is not in $F$, and every other edge of $\pi_i$ has its tail outside $X$ and thus is not in $F$ either because $\tail(F) \subseteq X$.
    Therefore, the concatenation of $\pi_1, \dots, \pi_k$ is a walk from $x$ to $y$ in $G \setminus F$ of the same weight, which is at least $\dist_{G \setminus F}(x, y)$.
\end{proof}

\paragraph*{$f$-sparsified shortcut graphs.}

The shortcut graph $\mathcal{S}(X)$ has at most $n|X|$ vertices and $2n|X|^2$ edges.
When we consider at most $f$ edge failures, a statement similar to \cref{lem:preserve} holds even for a sparser subgraph of $\mathcal{S}(X)$, which we define as follows.

\begin{definition}[$f$-sparsified shortcut graph]
    For each $x, y \in X$, let $A_X(x, y)$ be the set of shortcut edges $(e, y)$ of $\mathcal{S}(X)$ with $e \in \Eout(x)$ and $\delta_X(e, y) < \infty$.
    For each $x, y \in X$, let $L_X(x, y) \subseteq A_X(x, y)$ be a set satisfying the following conditions.
    \begin{itemize}
        \item $|L_X(x, y)| = \min\{f+1, |A_X(x, y)|\}$.
        \item $\delta_X(e, y) \le \delta_X(e', y)$ for every $(e, y) \in L_X(x, y)$ and every $(e', y) \in A_X(x, y) \setminus L_X(x, y)$.
    \end{itemize}
    An $f$-sparsified shortcut graph $\mathcal{S}_f(X)$ is the subgraph of $\mathcal{S}(X)$ that consists of the vertices of $X$, the shortcut edges in $L_X(x, y)$ for all $x, y \in X$, and, for each such shortcut edge $(e, y)$, the vertex $e$ and the departure edge $(\tail(e), e)$.
\end{definition}

In other words, $L_X(x, y)$ consists of $\min\{f+1, |A_X(x, y)|\}$ shortcut edges $(e, y) \in A_X(x, y)$ with the smallest values of $\delta_X(e, y)$, where ties are broken arbitrarily.
When at most $f$ edges fail, $\mathcal{S}_f(X)$ serves as a substitute for $\mathcal{S}(X)$, as shown in the following counterpart of \cref{lem:preserve}.

\begin{lemma}\label{lem:sparsify}
    Let $X \subseteq V$, let $x, y \in X$ and let $F \subseteq E$ be an edge set with $\tail(F) \subseteq X$ and $|F| \le f$.
    Then, $\dist_{G \setminus F}(x, y) = \dist_{\mathcal{S}(X) \setminus F}(x, y) = \dist_{\mathcal{S}_f(X) \setminus F}(x, y)$ holds.
\end{lemma}

\begin{proof}
    The first equality is \cref{lem:preserve}.
    Since $\mathcal{S}_f(X) \setminus F$ is a subgraph of $\mathcal{S}(X) \setminus F$, we have $\dist_{\mathcal{S}(X) \setminus F}(x, y) \le \dist_{\mathcal{S}_f(X) \setminus F}(x, y)$.
    For the converse, let $P$ be a shortest path from $x$ to $y$ in $\mathcal{S}(X) \setminus F$, and suppose that $P$ uses a shortcut edge $(e, v)$ that is not in $\mathcal{S}_f(X)$.
    We show that the subpath of $P$ from $\tail(e)$ to $v$ through $e$ can be replaced by a path in $\mathcal{S}_f(X) \setminus F$ of no larger weight.
    Let $u = \tail(e)$.
    Since the only incoming edge of $e$ in $\mathcal{S}(X)$ is $(u, e)$, the path $P$ contains the subpath $\langle u, e, v \rangle$.
    As the edge $(e, v)$ belongs to $A_X(u, v) \setminus L_X(u, v)$, $L_X(u, v)$ consists of $f+1$ edges, none of which is heavier than $\delta_X(e, v)$.
    Because $f+1$ edges in $L_X(u, v)$ have distinct tails and $|F| \le f$, at least one edge of $L_X(u, v)$, say $(e', v)$, satisfies $e' \notin F$.
    Thus, the path $\langle u, e', v \rangle$ exists in $\mathcal{S}_f(X) \setminus F$, and its weight is at most that of the path $\langle u, e, v \rangle$.
    Replacing $\langle u, e, v \rangle$ by $\langle u, e', v \rangle$ in $P$ never increases its weight.
\end{proof}

By definition, $\mathcal{S}_f(X)$ has at most $(f+1)\cdot|X|\cdot(|X|-1)$ shortcut edges, and hence $O(f|X|^2)$ vertices and edges.
Therefore, by \cref{lem:sparsify}, we can compute $\dist_{G \setminus F}(x, y)$ simply by computing the distance in $\mathcal{S}_f(X) \setminus F$.

\begin{corollary}\label{cor:query-alg}
    Let $X \subseteq V$.
    Given $\mathcal{S}_f(X)$, two vertices $x, y \in X$, and an edge set $F$ with $\tail(F) \subseteq X$ and $|F| \le f$, the distance $\dist_{G \setminus F}(x, y)$ can be computed in $O(f^2|X|^4)$ time.
\end{corollary}

\begin{proof}
    Let $H = \mathcal{S}_f(X) \setminus F$.
    Since $G$ has no negative cycle, neither does $H$.
    The graph $H$ has $O(f|X|^2)$ vertices and edges.
    Thus, the Bellman--Ford algorithm computes $\dist_H(x,y)$
    in $O(f^2|X|^4)$ time.
    By \cref{lem:sparsify}, this distance equals $\dist_{G\setminus F}(x,y)$.
\end{proof}

\subsection{Removal Algorithm} \label{subsec:removal}

In this subsection, we give an algorithm that computes $\mathcal{S}_f(X \setminus x)$ from $\mathcal{S}_f(X)$ for a vertex $x \in X$.
The algorithm is a key component in the next subsection, where we compute all the $f$-sparsified shortcut graphs needed to obtain $\dist_{G \setminus F}(s, t)$ for every failed edge set $F \subseteq E$ with $|F| \le f$.
The pseudocode is given in \cref{alg:remove}.

We represent $\mathcal{S}_f(X)$ by the sets $L_X(u, v)$ for all $u, v \in X$, where each shortcut edge $(e, v)$ in $L_X(u, v)$ is stored together with its weight $\delta_X(e, v)$.
In \cref{alg:remove}, the set $\mathcal{C}$ may contain two copies of the same shortcut edge $(e, v)$ with different weights.

\begin{algorithm}[h]
    \caption{$\textsc{Remove}(\mathcal{S}_f(X), x)$} \label{alg:remove}
    \KwIn{the lists $L_X(\ast, \ast)$ representing $\mathcal{S}_f(X)$, and a vertex $x \in X$}
    \KwOut{the lists $L_{X \setminus x}(\ast, \ast)$ representing $\mathcal{S}_f(X \setminus x)$}
    \BlankLine
    \ForEach{two vertices $u, v$ in $X \setminus x$}{
        $c \gets \min\{ \delta_X(e', v) \mid (e', v) \in L_X(x, v) \}$\;
        $\mathcal{C} \gets \{ (e, v) \text{ of weight } \delta_X(e, v) \mid (e, v) \in L_X(u, v) \}$ \;
        \If{$c < \infty$}{
            add $(e, v)$ of weight $\delta_X(e, x) + c$ to $\mathcal{C}$ for each $(e, x) \in L_X(u, x)$ \;
        }
        \ForEach{shortcut edge $(e, v)$ that appears twice in $\mathcal{C}$}{
            keep only the lighter copy of $(e, v)$ in $\mathcal{C}$ \label{line:keep}\;
        }
        $L_{X \setminus x}(u, v) \gets$ the $\min\{f + 1, |\mathcal{C}|\}$ lightest edges of $\mathcal{C}$ \label{line:select}\;
    }
    \BlankLine
    \Return{$L_{X \setminus x}$}\;
\end{algorithm}

\Cref{alg:remove} is based on the recurrence in \cref{lem:removal}.
When $x$ is removed from $X$, shortcuts are allowed to pass through $x$.
Hence, an $(X \setminus x)$-shortcut from $u$ to $v$ with first edge $e$ is either an $X$-shortcut from $u$ to $v$, or the concatenation of an $X$-shortcut from $u$ to $x$ with first edge $e$ and an $X$-shortcut from $x$ to $v$.
In the latter case, we may use a lightest $X$-shortcut from $x$ to $v$, whose weight is $c = \min_{e' \in \Eout(x)} \delta_X(e', v)$.
Moreover, we can compute $\mathcal{S}_f(X \setminus x)$ only from the shortcut edges stored in $\mathcal{S}_f(X)$, because $\mathcal{S}_f(X \setminus x)$ keeps only $f+1$ lightest shortcut edges for each pair of vertices.
Intuitively, if a shortcut edge is not stored in $L_X(u, v)$ or $L_X(u, x)$, then the same list stores $f+1$ other shortcut edges that are no heavier, and hence the shortcut edge is not needed for the $f+1$ lightest ones.



\begin{lemma}\label{lem:remove-time}
    Let $X \subseteq V$ and $x \in X$.
    Given $\mathcal{S}_f(X)$, \cref{alg:remove} computes $\mathcal{S}_f(X \setminus x)$ in $O(f^2 \cdot |X|^2)$ time.
\end{lemma}

\begin{proof}
    Fix $u, v \in X \setminus x$.
    Since $L_X(x, v)$ contains a lightest edge of $A_X(x, v)$ unless $A_X(x, v) = \emptyset$, we have $c = \min_{e' \in \Eout(x)} \delta_X(e', v)$.
    Hence, by \cref{lem:removal}, each shortcut edge $(e, v)$ of $\mathcal{S}(X \setminus x)$ has weight $\delta_{X \setminus x}(e, v) = \min\{\delta_X(e, v), \delta_X(e, x) + c\}$.
    After the lighter copies are kept, the weight of each edge $(e, v)$ in $\mathcal{C}$ is the minimum of the terms stored in the lists, that is, $\delta_X(e, v)$ if $(e, v) \in L_X(u, v)$ and $\delta_X(e, x) + c$ if $(e, x) \in L_X(u, x)$.
    If the smaller of the two terms $\delta_X(e, v)$ and $\delta_X(e, x) + c$ is stored, then the weight equals $\delta_{X \setminus x}(e, v)$.
    Otherwise, only the larger term is stored, and the weight can be larger than $\delta_{X \setminus x}(e, v)$.

    We next show that such a larger weight causes no problem, because a term is not stored only if the list contains $f+1$ other edges that are no heavier.
    Indeed, suppose that $\delta_{X \setminus x}(e, v) = \delta_X(e, v)$ and $(e, v) \notin L_X(u, v)$.
    By the definition of $L_X(u, v)$, the list $L_X(u, v)$ then consists of $f+1$ edges $(e', v)$ with $e' \neq e$ and $\delta_X(e', v) \le \delta_X(e, v)$, and all of them are in $\mathcal{C}$ with weight at most $\delta_{X \setminus x}(e, v)$.
    The case where $\delta_{X \setminus x}(e, v) = \delta_X(e, x) + c$ and $(e, x) \notin L_X(u, x)$ is analogous with $L_X(u, x)$ in place of $L_X(u, v)$.
    Therefore, every shortcut edge $(e, v) \in A_{X \setminus x}(u, v)$ satisfies one of the following.
    \begin{enumerate}[label=(\alph*)]
        \item The edge $(e, v)$ appears in $\mathcal{C}$, and its weight in $\mathcal{C}$ is $\delta_{X \setminus x}(e, v)$.
        \item The set $\mathcal{C}$ contains $f+1$ edges other than $(e, v)$ with weight at most $\delta_{X \setminus x}(e, v)$.
    \end{enumerate}

    We now show that the selected edges satisfy the following three properties, which are the conditions on $L_{X \setminus x}(u, v)$ in the definition of $\mathcal{S}_f(X \setminus x)$.
    \begin{enumerate}[label=(\roman*)]
        \item Every selected edge $(e, v)$ belongs to $A_{X \setminus x}(u, v)$ and has weight $\delta_{X \setminus x}(e, v)$ in $\mathcal{C}$.
        \item $\delta_{X \setminus x}(e', v) \le \delta_{X \setminus x}(e, v)$ for every selected edge $(e', v)$ and every unselected edge $(e, v) \in A_{X \setminus x}(u, v)$.
        \item The number of selected edges is $\min\{f+1, |A_{X \setminus x}(u, v)|\}$.
    \end{enumerate}
    Every edge $(e, v)$ in $\mathcal{C}$ has finite weight, which is at least $\delta_{X \setminus x}(e, v)$, and hence $\mathcal{C}$ contains only edges of $A_{X \setminus x}(u, v)$.

    We first show (i).
    Every selected edge belongs to $A_{X \setminus x}(u, v)$ by the observation above, and its weight in $\mathcal{C}$ is at least $\delta_{X \setminus x}(e, v)$.
    Hence, it suffices to show that no edge $(e, v)$ whose weight in $\mathcal{C}$ is larger than $\delta_{X \setminus x}(e, v)$ is selected.
    Let $(e, v)$ be such an edge.
    Then, (a) does not hold for $(e, v)$, and hence (b) holds.
    Thus, $\mathcal{C}$ contains $f+1$ edges whose weights are at most $\delta_{X \setminus x}(e, v)$, which is smaller than the weight of $(e, v)$ in $\mathcal{C}$.
    Therefore, $(e, v)$ is not among the $\min\{f+1, |\mathcal{C}|\}$ lightest edges of $\mathcal{C}$, regardless of how ties are broken.

    We next show (ii).
    Let $(e, v) \in A_{X \setminus x}(u, v)$ be an edge that is not selected.
    If (a) holds, then every selected edge has weight at most $\delta_{X \setminus x}(e, v)$ in $\mathcal{C}$, because the algorithm selects the lightest edges of $\mathcal{C}$.
    If (b) holds, then $\mathcal{C}$ contains $f+1$ edges of weight at most $\delta_{X \setminus x}(e, v)$, and hence every selected edge also has weight at most $\delta_{X \setminus x}(e, v)$ in $\mathcal{C}$.
    Since every selected edge $(e', v)$ has weight $\delta_{X \setminus x}(e', v)$ in $\mathcal{C}$ by (i), we obtain $\delta_{X \setminus x}(e', v) \le \delta_{X \setminus x}(e, v)$.

    Finally, we show (iii).
    Since each edge appears in $\mathcal{C}$ at most once, we have $|\mathcal{C}| \le |A_{X \setminus x}(u, v)|$.
    If $|\mathcal{C}| \ge f+1$, the algorithm selects $f+1$ edges, and $|A_{X \setminus x}(u, v)| \ge f+1$.
    If $|\mathcal{C}| < f+1$, then (b) never holds, and $\mathcal{C}$ contains all edges of $A_{X \setminus x}(u, v)$ by (a).
    Thus, the algorithm selects $\min\{f+1, |A_{X \setminus x}(u, v)|\}$ edges.

    By (i)--(iii), the algorithm correctly computes $L_{X \setminus x}(u, v)$ for all $u, v \in X \setminus x$, and hence it outputs $\mathcal{S}_f(X \setminus x)$.

    We now bound the running time.
    For each choice of $u$ and $v$, since each list $L_X(\ast, \ast)$ has at most $f+1$ edges, we have $|\mathcal{C}| = O(f)$, and computing $c$ and constructing $\mathcal{C}$ take $O(f)$ time.
    We keep the lighter copies by comparing every two edges of $\mathcal{C}$, and we select the lightest edges by sorting $\mathcal{C}$ by weight, both in $O(f^2)$ time.
    Since there are $O(|X|^2)$ choices of $u$ and $v$, the total running time is $O(f^2 |X|^2)$.
\end{proof}

\subsection{Constructing All Small Shortcut Graphs} \label{subsec:build}

In this subsection, we construct a data structure that consists of all the $f$-sparsified shortcut graphs needed to compute $\dist_{G \setminus F}(s, t)$.
By \cref{cor:query-alg}, we can compute $\dist_{G \setminus F}(s, t)$ in $O(f^6)$ time from the $f$-sparsified shortcut graph of a vertex set of size at most $f+2$ that contains $s$, $t$, and $\tail(F)$.
Since $F$ varies over the queries, we compute the $f$-sparsified shortcut graphs of all such vertex sets in advance.
To handle FRP, FSSRP, and FAPRP in a unified way, we introduce a vertex set $R \subseteq V$, which we call the \emph{fixed root set}.
It consists of the vertices among $s$ and $t$ that are fixed by the input, i.e., $R = \{s, t\}$ for FRP, $R = \{s\}$ for FSSRP, and $R = \emptyset$ for FAPRP.
Our goal is to compute $\mathcal{S}_f(R \cup X)$ for all $X \subseteq V$ with $1 \le |X| \le f$.

\paragraph*{Blocks and states.}
Recall that $n = 2^h$ and $V = \{v_0, v_1, \dots, v_{n-1}\}$.
For $0 \le j \le h$, the \emph{level-$j$ blocks} are the $2^j$ disjoint sets 
\[
    \mathcal{B}_j = \{B_{j, 0}, \dots, B_{j, 2^j-1}\}, \qquad B_{j, i} = \{v_k \mid i\cdot 2^{h-j} \le k < (i+1) \cdot 2^{h-j} \} \qquad (0 \le i < 2^j),
\]
each of size $n/2^j$.
Thus, $\mathcal{B}_0 = \{V\}$, the set $\mathcal{B}_1$ consists of the two halves of $V$, and so on, until $\mathcal{B}_h = \left \{ \{v_0\}, \dots, \{v_{n-1}\} \right \}$.
Each block in $\mathcal{B}_{j+1}$ is one of the two halves of a block in $\mathcal{B}_j$.




To enumerate the vertex sets $R \cup X$ with $1 \le |X| \le f$, we use $f$-tuples of blocks.
A \emph{level-$j$ state} is an $f$-tuple $\mathbf{B} = (B_1, B_2, \dots, B_f) \in \mathcal{B}_j^f$ of level-$j$ blocks, which are not necessarily distinct, and it represents the vertex set $X(\mathbf{B}) = R \cup B_1 \cup \dots \cup B_f$.
Since every level-$h$ block consists of a single vertex, the vertex sets of the level-$h$ states are exactly the sets $R \cup X$ with $X \subseteq V$ and $1 \le |X| \le f$.
Since $|\mathcal{B}_j| = 2^j$ and each level-$j$ block has $n/2^j$ vertices, we have the following.
\begin{itemize}
    \item The number of level-$j$ states is $|\mathcal{B}_j^f| = 2^{jf}$.
    \item The vertex set of every level-$j$ state has at most $|R| + f \cdot n/2^j$ vertices.
\end{itemize}

\begin{figure}[!tb]
    \centering
    \includegraphics[width=\textwidth]{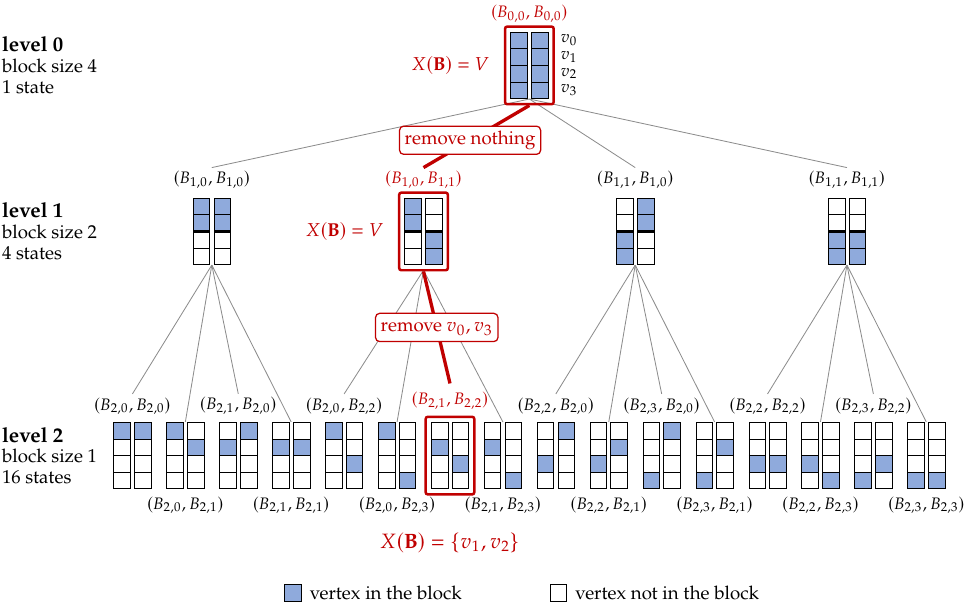}
    \caption{The tree of states for $n = 4$, $f = 2$, and $R = \emptyset$.
    The vertex set $X(\mathbf{B})$ consists of the vertices that are blue in at least one of the two columns.
    Each edge connects a state to its parent.
    The red path shows how \cref{alg:build} obtains $\mathcal{S}_2(\{v_1, v_2\})$ from $\mathcal{S}_2(V)$ by removing the indicated vertices with \cref{alg:remove}.}\label{fig:state-tree}
\end{figure}

The \emph{parent} of a level-$(j+1)$ state $\mathbf{B}' = (B_1', \dots, B_f')$ is the level-$j$ state $\mathbf{B} = (B_1, \dots, B_f)$, where $B_i$ is the level-$j$ block containing $B_i'$ for each $i$.
Since $B_i' \subseteq B_i$ for each $i$, we have $X(\mathbf{B}') \subseteq X(\mathbf{B})$.
Hence, the states form a tree rooted at the level-$0$ state $(V, \dots, V)$, and the vertex sets shrink from the root toward the leaves (see \cref{fig:state-tree}).

\paragraph*{Algorithm and analysis.}
We process the states level by level, from level $0$ to level $h$.
The only level-$0$ state is $(V, \dots, V)$, whose vertex set is $V$, and $\mathcal{S}_f(V)$ is easy to construct because every $V$-shortcut consists of a single edge.
For each $j = 0, 1, \dots, h-1$, we compute the $f$-sparsified shortcut graph of every level-$(j+1)$ state $\mathbf{B}'$ from that of its parent $\mathbf{B}$, which has already been computed.
Since $X(\mathbf{B}') \subseteq X(\mathbf{B})$, we obtain $\mathcal{S}_f(X(\mathbf{B}'))$ by applying \textsc{Remove} (\cref{alg:remove}) to a copy of $\mathcal{S}_f(X(\mathbf{B}))$ for the vertices in $X(\mathbf{B}) \setminus X(\mathbf{B}')$ one by one.
The details are given in \cref{alg:build}, and \cref{fig:state-tree} illustrates the process.

\begin{algorithm}[h]
    \caption{$\textsc{BuildAll}(G, R, f)$} \label{alg:build}
    \KwIn{a graph $G = (V, E, w)$ with $|V| = 2^h$, a fixed root set $R \subseteq V$, and an integer $f \ge 1$}
    \KwOut{$\mathcal{S}_f(R \cup X)$ for all $X \subseteq V$ with $1 \le |X| \le f$}
    \BlankLine
    \ForEach{two vertices $u, v$ in $V$}{
        $L_V(u, v) \gets \{ ((u, v), v) \}$ with weight $w(u, v)$ if $(u, v) \in E$, and $L_V(u, v) \gets \emptyset$ otherwise\;
    }
    store the lists $L_V(\ast, \ast)$ as $\mathcal{S}_f(V)$, which is $\mathcal{S}_f(X(\mathbf{B}))$ for the level-$0$ state $\mathbf{B} = (V, \dots, V)$\;
    \For{$j = 0, 1, \dots, h - 1$}{
        \ForEach{level-$(j+1)$ state $\mathbf{B}'$}{
            $\mathbf{B} \gets$ the parent of $\mathbf{B}'$\;
            $H \gets$ a copy of $\mathcal{S}_f(X(\mathbf{B}))$\;
            \ForEach{$x \in X(\mathbf{B}) \setminus X(\mathbf{B}')$}{
                $H \gets \textsc{Remove}(H, x)$\;
            }
            store $H$ as $\mathcal{S}_f(X(\mathbf{B}'))$\;
        }
    }
    \Return{$\mathcal{S}_f(X(\mathbf{B}))$ for all level-$h$ states $\mathbf{B}$}\;
\end{algorithm}


\begin{lemma}\label{lem:build-analysis}
    Let $R \subseteq V$.
    \cref{alg:build} computes $\mathcal{S}_f(R \cup X)$ for all $X \subseteq V$ with $1 \le |X| \le f$ in $O(2^f f^2 (f + |R|)^3 \cdot T_f (n))$ time, where
    \begin{equation} \label{eq:time}
        T_f(n) =
        \begin{cases}
            n^3 & \text{if } f \le 2, \\
            n^3 \log n & \text{if } f = 3, \\
            n^f & \text{if } f \ge 4.
        \end{cases}
    \end{equation}
\end{lemma}

\begin{proof}
    We first show the correctness.
    Since every $V$-shortcut consists of a single edge, the initialization correctly computes $\mathcal{S}_f(V)$, which is the graph of the level-$0$ state.
    By \cref{lem:remove-time}, each call of \cref{alg:remove} transforms $\mathcal{S}_f(Y)$ into $\mathcal{S}_f(Y \setminus x)$ for the current vertex set $Y$ and $x \in Y$.
    Hence, by induction on $j$, the algorithm stores $\mathcal{S}_f(X(\mathbf{B}))$ for every level-$j$ state $\mathbf{B}$.
    Since the vertex sets of the level-$h$ states are exactly the sets $R \cup X$ with $X \subseteq V$ and $1 \le |X| \le f$, the algorithm outputs $\mathcal{S}_f(R \cup X)$ for all such $X$.

    We next bound the running time.
    The initialization takes $O(n^2)$ time.
    Let $n_j = (|R| + f)\cdot n/2^j$.
    Recall that there are $2^{jf}$ level-$j$ states, and the vertex set of each of them has at most $|R| + f \cdot n/2^j \le n_j$ vertices.
    For each level-$(j+1)$ state with parent $\mathbf{B}$, the algorithm copies the lists representing $\mathcal{S}_f(X(\mathbf{B}))$, which contain $O(fn_j^2)$ edges, and calls \cref{alg:remove} at most $|X(\mathbf{B})| \le n_j$ times, each of which takes $O(f^2n_j^2)$ time by \cref{lem:remove-time}.
    Thus, each level-$(j+1)$ state is processed in $O(f^2n_j^3)$ time, and since there are $2^{(j+1)f}$ level-$(j+1)$ states, the total running time is
    \[
        O(n^2) + \sum_{j=0}^{h-1} 2^{(j+1)f} \cdot O(f^2n_j^3) = O \left (n^2 + 2^f f^2 (f+|R|)^3 \cdot n^3 \sum_{j=0}^{h-1}2^{(f-3)j} \right ).
    \]
    The sum is at most $2$ if $f \le 2$, it is $h = \log n$ if $f = 3$, and it is at most $2^{(f-3)h} = n^{f-3}$ if $f \ge 4$.
    In all cases, the running time is $O(2^f f^2 (f + |R|)^3 \cdot T_f (n))$.
\end{proof}

\subsection{Proof of Main Theorems} \label{subsec:proof}

In this subsection, we prove our main theorems, \cref{thm:2frp,thm:3frp,thm:4frp} and \cref{cor:ffrp}.
We first state our result in the form of a distance oracle, which answers $\dist_{G \setminus F}(s, t)$ for a given failed edge set $F$ after preprocessing.
The main theorems then follow by querying the oracle for all the outputs.

\begin{lemma}\label{lem:oracle}
    Let $f \ge 2$ be a fixed integer, and let $G = (V, E, w)$ be a directed graph with $n$ vertices, real edge weights, and no negative cycles.
    For each of the following three query types, there is an algorithm that, in $O(2^f f^5 \cdot T_f(n))$ preprocessing time, constructs a data structure that answers every query of that type in $O(f^6)$ time, where $T_f(n)$ is defined in \eqref{eq:time}.

    \begin{enumerate}[label=\textup{(\roman*)}]
        \item A source $s \in V$ and a target $t \in V$ are specified at preprocessing. A query is a set $F \subseteq E$ with $|F| \le f$. The answer is $\dist_{G \setminus F}(s, t)$.
        \item A source $s \in V$ is specified at preprocessing. A query is a pair $(t, F)$ with $t \in V$, $F \subseteq E$, and $|F| \le f-1$. The answer is $\dist_{G \setminus F}(s, t)$.
        \item A query is a triple $(s, t, F)$ with $s, t \in V$, $F \subseteq E$, and $|F| \le f-2$. The answer is $\dist_{G \setminus F}(s, t)$.
    \end{enumerate}
\end{lemma}

\begin{proof}
    In all three cases, we run \cref{alg:build} and compute sparsified shortcut graphs of all level-$h$ states.
    The root set is $R = \{s, t\}$ in case (i), $R = \{s\}$ in case (ii), and $R = \emptyset$ in case (iii).
    Given a query, let $X = \tail(F)$ in case (i), $X = \tail(F) \cup \{t\}$ in case (ii), and $X = \tail(F) \cup \{s, t\}$ in case (iii).
    By \cref{cor:query-alg}, we can obtain $\dist_{G \setminus F}(s, t)$ in $O(f^6)$ time.
    Since $|R| \le 2$, the preprocessing time is $O(2^f f^5 \cdot T_f(n))$ by \cref{lem:build-analysis}.
\end{proof}

To output the distances explicitly, we enumerate the $O(n^f)$ edge sets $F$ described in \cref{sec:intro}, in which the $i$-th failed edge is chosen from a shortest $s$-$t$ path after the first $i-1$ failures.
Hence, for each such edge set $F$ with $|F| < f$, we need a shortest $s$-$t$ path in $G \setminus F$, not only its distance.
To obtain such a path, \cref{alg:remove} stores, for each shortcut edge of weight $\delta_X(e, x) + c$, pointers to the two shortcut edges $(e, x)$ and $(e', v)$ that give the weight, and reuses the other shortcut edges as they are.
By following these pointers recursively, we expand each shortcut edge on a shortest path in $\mathcal{S}_f(R \cup X) \setminus F$ into a path in $G$, which yields a shortest path in $G \setminus F$.
Since each pointer splits a shortcut at one internal vertex, the expansion takes $O(n)$ time.
Since we report a path only for $O(n^{f-1})$ edge sets $F$, the total time for reporting paths is $O(n^f)$.

Finally, we show our main theorems.

\begin{proof}[Proof of \cref{thm:2frp,thm:3frp,thm:4frp}]
    Let $f \in \{2, 3, 4\}$.
    Since $f$ is a constant, the oracle of \cref{lem:oracle} is constructed in $O(T_f(n))$ time, answers a query in $O(1)$ time, and reports a shortest path in $O(n)$ time, where we use type (i) for $f$-FRP, type (ii) for $(f-1)$-FSSRP, and type (iii) for $(f-2)$-FAPRP.
    It then suffices to query the oracle for each of the $O(n^f)$ outputs, which takes $O(n^f)$ time in total.
    Hence, the running time is $O(T_f(n))$.
\end{proof}

\begin{proof}[Proof of \cref{cor:ffrp}]
    We proceed by induction on $f$.
    The base case $f = 4$ is \cref{thm:4frp}, so let $f \ge 5$.

    For $f$-FRP, we compute a shortest $s$-$t$ path $\pi$ in $G$ by the Bellman--Ford algorithm in $O(n^3)$ time and, for each of the at most $n-1$ edges $e$ on $\pi$, solve $(f-1)$-FRP on $G \setminus e$.
    By the induction, this takes $O(n) \cdot O(n^{f-1}) = O(n^f)$ time.

    For $(f-1)$-FSSRP, we solve $(f-1)$-FRP for each of the $n$ targets.
    Since $(f-1)$-FRP is solved in $O(n^{f-1})$ time, this takes $O(n^f)$ time.

    For $(f-2)$-FAPRP, we solve $(f-2)$-FSSRP from each of the $n$ sources.
    Since $(f-2)$-FSSRP is solved in $O(n^{f-1})$ time, this takes $O(n^f)$ time.
\end{proof}

\section{Further Results}
Our approach extends in several directions.
We only state the results here and defer the details to a forthcoming manuscript.\footnote{We are well aware that stating the further results in this section without proofs is not an ideal practice.
However, the current write-ups of these results are drafts prepared with the assistance of AI tools, and we do not consider it appropriate to include such drafts as they are.}
\begin{itemize}
    \item \textbf{$2$-FRP in APSP time.}
    If APSP in directed weighted graphs can be solved in $O(n^\alpha)$ time for some $\alpha > 2$, then $2$-FRP in directed weighted graphs can be solved in $O(n^\alpha)$ time.
    \item \textbf{$3$-FRP in cubic time.}
    $3$-FRP in directed weighted graphs can be solved in deterministic $O(n^3)$ time, which removes the logarithmic factor in \cref{thm:3frp}.
    The key idea is to remove many vertices at once by a single APSP computation.
\end{itemize}

\paragraph{Acknowledgements.}
Kaito Harada was supported by JST BOOST, Japan Grant Number JPMJBS2402.

\paragraph{AI Disclosure.}
The main ideas behind our algorithm based on shortcut graphs were developed largely with the assistance of GPT-6 Astra (OpenAI).
We also used GPT-6 Astra (OpenAI), Claude Opus 5.5, and Claude Fable 5.1 (Anthropic) to assist in writing the proofs, preparing the figures, and improving the language and presentation.
All proofs and the text of this paper were verified and revised by the authors, who take full responsibility for its content.

\bibliographystyle{alphaurl}
\bibliography{biblio}

@article{malik1989k,
  author  = {Kavindra Malik and Ashok K. Mittal and Santosh K. Gupta},
  title   = {The $k$ Most Vital Arcs in the Shortest Path Problem},
  journal = {Operations Research Letters},
  volume  = {8},
  number  = {4},
  pages   = {223--227},
  year    = {1989},
  doi     = {10.1016/0167-6377(89)90065-5}
}

@article{nardelli2001faster,
  author  = {Enrico Nardelli and Guido Proietti and Peter Widmayer},
  title   = {A Faster Computation of the Most Vital Edge of a Shortest Path},
  journal = {Information Processing Letters},
  volume  = {79},
  number  = {2},
  pages   = {81--85},
  year    = {2001},
  doi     = {10.1016/S0020-0190(00)00175-7}
}

@inproceedings{bhosle2004replacement,
  author    = {Amit M. Bhosle and Teofilo F. Gonzalez},
  title     = {Replacement Paths for Pairs of Shortest Path Edges in Directed Graphs},
  booktitle = {Proc. of the 16th {IASTED} International Conference on Parallel and Distributed Computing and Systems (PDCS)},
  year      = {2004}
}

@article{hershberger2007difficulty,
  author  = {John Hershberger and Subhash Suri and Amit Bhosle},
  title   = {On the Difficulty of Some Shortest Path Problems},
  journal = {ACM Transactions on Algorithms},
  volume  = {3},
  number  = {1},
  pages   = {5:1--5:15},
  year    = {2007},
  doi     = {10.1145/1186810.1186815}
}

@inproceedings{BK09,
  author    = {Aaron Bernstein and David R. Karger},
  title     = {A Nearly Optimal Oracle for Avoiding Failed Vertices and Edges},
  booktitle = {Proc. of the 41st Annual {ACM} Symposium on Theory of Computing (STOC)},
  pages     = {101--110},
  year      = {2009},
  doi       = {10.1145/1536414.1536431}
}

@article{gotthilf2009improved,
  author  = {Zvi Gotthilf and Moshe Lewenstein},
  title   = {Improved Algorithms for the $k$ Simple Shortest Paths and the Replacement Paths Problems},
  journal = {Information Processing Letters},
  volume  = {109},
  number  = {7},
  pages   = {352--355},
  year    = {2009},
  doi     = {10.1016/j.ipl.2008.12.015}
}

@article{emek2010near,
  author  = {Yuval Emek and David Peleg and Liam Roditty},
  title   = {A Near-Linear-Time Algorithm for Computing Replacement Paths in Planar Directed Graphs},
  journal = {ACM Transactions on Algorithms},
  volume  = {6},
  number  = {4},
  pages   = {64:1--64:13},
  year    = {2010},
  doi     = {10.1145/1824777.1824784}
}

@inproceedings{wulff2010solving,
  author    = {Christian Wulff-Nilsen},
  title     = {Solving the Replacement Paths Problem for Planar Directed Graphs in {$O(n \log n)$} Time},
  booktitle = {Proc. of the 21st Annual {ACM-SIAM} Symposium on Discrete Algorithms (SODA)},
  pages     = {756--765},
  year      = {2010},
  doi       = {10.1137/1.9781611973075.62}
}

@inproceedings{bernstein2010nearly,
  author    = {Aaron Bernstein},
  title     = {A Nearly Optimal Algorithm for Approximating Replacement Paths and $k$ Shortest Simple Paths in General Graphs},
  booktitle = {Proc. of the 21st Annual {ACM-SIAM} Symposium on Discrete Algorithms (SODA)},
  pages     = {742--755},
  year      = {2010},
  doi       = {10.1137/1.9781611973075.61}
}

@inproceedings{erickson2011computing,
  author    = {Jeff Erickson and Amir Nayyeri},
  title     = {Computing Replacement Paths in Surface Embedded Graphs},
  booktitle = {Proc. of the 22nd Annual {ACM-SIAM} Symposium on Discrete Algorithms (SODA)},
  pages     = {1347--1354},
  year      = {2011},
  doi       = {10.1137/1.9781611973082.103}
}

@inproceedings{williams2011faster,
  author    = {Virginia {Vassilevska Williams}},
  title     = {Faster Replacement Paths},
  booktitle = {Proc. of the 22nd Annual {ACM-SIAM} Symposium on Discrete Algorithms (SODA)},
  pages     = {1337--1346},
  year      = {2011},
  doi       = {10.1137/1.9781611973082.102}
}

@article{RZ12,
  author  = {Liam Roditty and Uri Zwick},
  title   = {Replacement Paths and $k$ Simple Shortest Paths in Unweighted Directed Graphs},
  journal = {ACM Transactions on Algorithms},
  volume  = {8},
  number  = {4},
  pages   = {33:1--33:11},
  year    = {2012},
  doi     = {10.1145/2344422.2344423}
}

@inproceedings{GV12,
  author    = {Fabrizio Grandoni and Virginia {Vassilevska Williams}},
  title     = {Improved Distance Sensitivity Oracles via Fast Single-Source Replacement Paths},
  booktitle = {Proc. of the 53rd Annual {IEEE} Symposium on Foundations of Computer Science (FOCS)},
  pages     = {748--757},
  year      = {2012},
  doi       = {10.1109/FOCS.2012.17}
}

@article{weimann2013replacement,
  author  = {Oren Weimann and Raphael Yuster},
  title   = {Replacement Paths and Distance Sensitivity Oracles via Fast Matrix Multiplication},
  journal = {ACM Transactions on Algorithms},
  volume  = {9},
  number  = {2},
  pages   = {14:1--14:13},
  year    = {2013},
  doi     = {10.1145/2438645.2438646}
}

@article{lee2014replacement,
  author  = {Cheng-Wei Lee and Hsueh-I Lu},
  title   = {Replacement Paths via Row Minima of Concise Matrices},
  journal = {SIAM Journal on Discrete Mathematics},
  volume  = {28},
  number  = {1},
  pages   = {206--225},
  year    = {2014},
  doi     = {10.1137/120897146}
}

@inproceedings{ACC19,
  author    = {Noga Alon and Shiri Chechik and Sarel Cohen},
  title     = {Deterministic Combinatorial Replacement Paths and Distance Sensitivity Oracles},
  booktitle = {Proc. of the 46th International Colloquium on Automata, Languages, and Programming (ICALP)},
  pages     = {12:1--12:14},
  year      = {2019},
  doi       = {10.4230/LIPIcs.ICALP.2019.12}
}

@inproceedings{CC19,
  author    = {Shiri Chechik and Sarel Cohen},
  title     = {Near Optimal Algorithms for the Single Source Replacement Paths Problem},
  booktitle = {Proc. of the 30th Annual {ACM-SIAM} Symposium on Discrete Algorithms (SODA)},
  pages     = {2090--2109},
  year      = {2019},
  doi       = {10.1137/1.9781611975482.126}
}

@article{grandoni2019faster,
  author  = {Fabrizio Grandoni and Virginia {Vassilevska Williams}},
  title   = {Faster Replacement Paths and Distance Sensitivity Oracles},
  journal = {ACM Transactions on Algorithms},
  volume  = {16},
  number  = {1},
  pages   = {15:1--15:25},
  year    = {2019},
  doi     = {10.1145/3365835}
}

@inproceedings{chechik2020simplifying,
  author    = {Shiri Chechik and Moran Nechushtan},
  title     = {Simplifying and Unifying Replacement Paths Algorithms in Weighted Directed Graphs},
  booktitle = {Proc. of the 47th International Colloquium on Automata, Languages, and Programming (ICALP)},
  pages     = {29:1--29:12},
  year      = {2020},
  doi       = {10.4230/LIPIcs.ICALP.2020.29}
}

@inproceedings{CM20,
  author    = {Shiri Chechik and Ofer Magen},
  title     = {Near Optimal Algorithm for the Directed Single Source Replacement Paths Problem},
  booktitle = {Proc. of the 47th International Colloquium on Automata, Languages, and Programming (ICALP)},
  pages     = {81:1--81:17},
  year      = {2020},
  doi       = {10.4230/LIPIcs.ICALP.2020.81}
}

@inproceedings{GJM20,
  author    = {Manoj Gupta and Rahul Jain and Nitiksha Modi},
  title     = {Multiple Source Replacement Path Problem},
  booktitle = {Proc. of the 39th {ACM} Symposium on Principles of Distributed Computing (PODC)},
  pages     = {339--348},
  year      = {2020},
  doi       = {10.1145/3382734.3405714}
}

@inproceedings{BCFS21,
  author    = {Davide Bil{\`o} and Sarel Cohen and Tobias Friedrich and Martin Schirneck},
  title     = {Near-Optimal Deterministic Single-Source Distance Sensitivity Oracles},
  booktitle = {Proc. of the 29th Annual European Symposium on Algorithms (ESA)},
  pages     = {18:1--18:17},
  year      = {2021},
  doi       = {10.4230/LIPIcs.ESA.2021.18}
}

@inproceedings{GPVX21,
  author    = {Yuzhou Gu and Adam Polak and Virginia {Vassilevska Williams} and Yinzhan Xu},
  title     = {Faster Monotone Min-Plus Product, Range Mode, and Single Source Replacement Paths},
  booktitle = {Proc. of the 48th International Colloquium on Automata, Languages, and Programming (ICALP)},
  pages     = {75:1--75:20},
  year      = {2021},
  doi       = {10.4230/LIPIcs.ICALP.2021.75}
}

@inproceedings{BCCFS22,
  author    = {Davide Bil{\`o} and Keerti Choudhary and Sarel Cohen and Tobias Friedrich and Martin Schirneck},
  title     = {Deterministic Sensitivity Oracles for Diameter, Eccentricities and All Pairs Distances},
  booktitle = {Proc. of the 49th International Colloquium on Automata, Languages, and Programming (ICALP)},
  pages     = {22:1--22:19},
  year      = {2022},
  doi       = {10.4230/LIPIcs.ICALP.2022.22}
}

@inproceedings{DG22,
  author    = {Dipan Dey and Manoj Gupta},
  title     = {Near Optimal Algorithm for Fault Tolerant Distance Oracle and Single Source Replacement Path Problem},
  booktitle = {Proc. of the 30th Annual European Symposium on Algorithms (ESA)},
  pages     = {42:1--42:18},
  year      = {2022},
  doi       = {10.4230/LIPIcs.ESA.2022.42}
}

@inproceedings{VWX22,
  author    = {Virginia {Vassilevska Williams} and Eyob Woldeghebriel and Yinzhan Xu},
  title     = {Algorithms and Lower Bounds for Replacement Paths under Multiple Edge Failure},
  booktitle = {Proc. of the 63rd Annual {IEEE} Symposium on Foundations of Computer Science (FOCS)},
  pages     = {907--918},
  year      = {2022},
  doi       = {10.1109/FOCS54457.2022.00090}
}

@inproceedings{CZ24,
  author    = {Shiri Chechik and Tianyi Zhang},
  title     = {Faster Algorithms for Dual-Failure Replacement Paths},
  booktitle = {Proc. of the 51st International Colloquium on Automata, Languages, and Programming (ICALP)},
  pages     = {41:1--41:20},
  year      = {2024},
  doi       = {10.4230/LIPIcs.ICALP.2024.41}
}

@inproceedings{manoharan2024computing,
  author    = {Vignesh Manoharan and Vijaya Ramachandran},
  title     = {Computing Replacement Paths in the {CONGEST} Model},
  booktitle = {Proc. of the 31st International Colloquium on Structural Information and Communication Complexity (SIROCCO)},
  pages     = {420--437},
  year      = {2024},
  doi       = {10.1007/978-3-031-60603-8_23}
}

@inproceedings{chang2025optimal,
  author    = {Yi-Jun Chang and Yanyu Chen and Dipan Dey and Gopinath Mishra and Hung Thuan Nguyen and Bryce Sanchez},
  title     = {Optimal Distributed Replacement Paths},
  booktitle = {Proc. of the 44th {ACM} Symposium on Principles of Distributed Computing (PODC)},
  pages     = {287--298},
  year      = {2025},
  doi       = {10.1145/3732772.3733513}
}

@inproceedings{CDWX25,
  author    = {Shucheng Chi and Ran Duan and Benyu Wang and Tianle Xie},
  title     = {Undirected 3-Fault Replacement Path in Nearly Cubic Time},
  booktitle = {Proc. of the 52nd International Colloquium on Automata, Languages, and Programming (ICALP)},
  pages     = {57:1--57:20},
  year      = {2025},
  doi       = {10.4230/LIPIcs.ICALP.2025.57}
}

@inproceedings{NV26,
  author    = {Jakob Nogler and Virginia {Vassilevska Williams}},
  title     = {Undirected Replacement Paths: Dual Fault Reduces to Single Source},
  booktitle = {Proc. of the 53rd International Colloquium on Automata, Languages, and Programming (ICALP)},
  pages     = {144:1--144:16},
  year      = {2026},
  doi       = {10.4230/LIPIcs.ICALP.2026.144}
}

@misc{HI26,
  author        = {Kaito Harada and Taisuke Izumi},
  title         = {A Simple Algorithm for the Directed Multiple Source Replacement Paths Problem},
  year          = {2026},
  note          = {To appear in SOSA 2027},
  archivePrefix = {arXiv},
  eprint        = {2608.16615},
  primaryClass  = {cs.DS}
}

@misc{HT26,
  author        = {Kaito Harada and Tatsuya Terao},
  title         = {Simplified and Faster Algorithms for the Directed Single/Dual-Failure Replacement Paths via Modulo-Layer Hitting Sets},
  year          = {2026},
  archivePrefix = {arXiv},
  eprint        = {2610.24817},
  primaryClass  = {cs.DS}
}

@misc{alman2026trulysubquadratic3sumtruly,
  author        = {Josh Alman and Virginia {Vassilevska Williams}},
  title         = {Truly Subquadratic {3SUM} and Truly Subcubic {APSP} via Triangles in Sparse Lopsided Graphs},
  year          = {2026},
  archivePrefix = {arXiv},
  eprint        = {2610.06783},
  primaryClass  = {cs.DS}
}

@misc{Maj26,
  author        = {Ruhollah Majdoddin},
  title         = {Thin Matrix Products up to {$D \le N^{0.172}$} and Lossless Reductions for {3SUM}, {E}xact {T}riangle and {APSP}, with {3SUM} over Abelian Groups and a Dichotomy for Coordinatewise Problems},
  year          = {2026},
  archivePrefix = {arXiv},
  eprint        = {2610.17420},
  primaryClass  = {cs.DS}
}

@misc{CIMY26,
  author        = {Timothy M. Chan and Reina Itakura and Zhiyuan Ma and Yuancheng Yu},
  title         = {{3SUM} Made Simpler (Without Matrix Multiplication), {$k$-SUM} Hypothesis Refuted for Odd $k$, and Real {3SUM}/{APSP} Derandomized},
  year          = {2026},
  archivePrefix = {arXiv},
  eprint        = {2610.42012},
  primaryClass  = {cs.DS}
}

\end{document}